\documentclass{mem}
\usepackage{natbib}\usepackage{txfonts}\usepackage{balance}
\usepackage{graphicx}
\usepackage[a4paper]{hyperref}
\idline{123}{123}
\begin{document}

\title{The case of the disappearing CN-strong AGB stars in Galactic globular
  clusters -- preliminary results}

\author{
S. W.\, Campbell\inst{1,2},
D.\, Yong\inst{3},
E. C.\, Wylie-de Boer\inst{3},
R. J.\, Stancliffe\inst{2},
J. C.\, Lattanzio\inst{2},
G. C.\, Angelou\inst{2},
F. Grundahl\inst{4}
\and
C. Sneden\inst{5}
        }

\offprints{S.W. Campbell}
 
\institute{
           Departament de F\'{i}sica i Enginyeria Nuclear, 
           Universitat Polit\`{e}cnica de Catalunya, EUETIB, 
           Carrer Comte d'Urgell 187, E-08036 Barcelona, Spain.\\
           \email{simon.w.campbell@upc.edu}
\and
           Centre for Stellar and Planetary Astrophysics, 
           Monash University, PO Box 28M, Victoria 3800, Australia.
\and
           Research School of Astronomy and Astrophysics, 
           Australian National University, 
           Weston, ACT 2611, Australia.
\and
           Department of Physics and Astronomy, 
           Aarhus University, Ny Munkegade, 8000 Aarhus C, Denmark.
\and
           Department of Astronomy and McDonald Observatory, 
           University of Texas, Austin, TX 78712, USA. 
}

\authorrunning{Campbell et al.}

\titlerunning{Abundances of AGB Stars in GCs}


\abstract{

A previously reported literature search suggested that the AGB stars in
Galactic globular clusters may be showing different distributions of
CN-strong and CN-weak stars as compared to their RGB stars. In most cases
the second giant branches of GCs appeared to be deficient in stars with
strong CN bands. However the sample sizes of AGB stars at that time
were too small to give a definitive picture. Thus an observing campaign
targeting GC AGB stars was proposed. We now have medium resolution spectral
observations of about 250 GC AGB stars across 9 globular clusters, obtained
with the 2dF/AAOmega instrument on the Anglo-Australian Telescope. In this
paper we report some preliminary results regarding the distributions of
CN-strong and CN-weak stars on the two giant branches of a selection of
globular clusters. We find that some GCs show a total lack of CN-strong
stars on the AGB, whilst some show a reduction in CN-strong stars as
compared to the RGB. Standard stellar evolution does not predict this
change in surface abundance between the two giant branches. We discuss some
possible causes of this unexpected phenomenon.

          } 

\maketitle{}


\section{Introduction}

Although Galactic globular clusters (GCs) are chemically homogeneous with
respect to Fe and most other heavy elements (see e.g. \citealt{KSL92}), it has
long been known that they show inhomogeneities in many lighter elements
(e.g. C, N, O, Mg, Al). These inhomogeneities are considered anomalous
because they are not seen in halo field stars of similar metallicity (see
e.g. \citealt{GSC00}).

One of the first inhomogeneities discovered was that of the molecule
cyanogen (CN, often used as a proxy for nitrogen). A picture of
`CN-bimodality' emerged in the early 1980s whereby there appears to be two
distinct chemical populations of stars in most, if not all, GCs. One
population is known as `CN-strong', the other `CN-weak' (we note that the
CN-weak population might be more informatively called `CN-normal' -- as
these stars show CN abundances similar to the Halo field stars).
Originally, observations of CN were mainly made in stars on the giant
branches but more recently there have been observations on the main
sequence (MS) and sub-giant branch (SGB) of some clusters (e.g.
\citealt{CCB98}). These observations show that there is little difference
in the bimodal CN pattern on the MS and SGB as compared with the giants ---
indicating a primordial origin for the differing populations. Figure 6 in
\citet{CCB98} exemplifies this situation.

Due to the paucity of asymptotic giant branch (AGB) stars in GCs (a result
of their short lifetimes) there have been very few systematic observational
studies of the CN anomaly on the AGB in globular clusters (\citealt{M78} is
one of which we are aware).  What little that has been done has
been an aside in more general papers (e.g. \citealt{NCF81}, \citealt{BSH93},
\citealt{ISK99}). However these studies have hinted at a tantalising
characteristic: most (observed) GCs show a lack of CN-strong stars on the
AGB. If this is true then it is in stark contrast to the red giant branch
(RGB) and earlier phases of evolution, where the ratio of CN-Strong to
CN-Weak stars is roughly unity in many clusters.

This \textit{possible} discrepancy was noted by \citet{NCF81} in their
paper about abundances in giant stars in NGC 6752. They state that ``The
behaviour of the CN bands in the AGB stars is... quite difficult to
understand... not one of the stars studied here has enhanced CN... yet on
the [first] giant branch there are more CN strong stars than CN weak
ones.''  (also see Figure 3 of that paper). 

Motivated by this observation we conducted a literature search to ascertain
whether this was true of other GCs (\citealt{Paper1}). The literature
search revealed that the AGB star counts for all studies (which are not, in
general, studies about AGB stars in particular) are low, usually being
$\leq 10$ (see Table 1 of \citealt{Paper1}). The search also revealed that
the picture may not be consistent between clusters. Although most clusters
appear to have CN-weak AGBs, at least two seem to have CN-strong AGBs (M5
\& 47 Tuc).  To further complicate the picture, clusters often appear to
have a combination of both CN-strong and CN-weak stars on their
AGBs. Again, all these assertions are however based on small sample sizes.

We note that \citet{SIK00} also presented a conference paper on this exact
topic. Compiling the data in the literature at the time they discussed the
relative amounts of CN in AGB and RGB stars in the GCs. They also discuss
Na abundance variations in M13.  In their closing remarks they suggest
observations with larger sample sizes are needed --- which may be done
using wide-field multi-object spectroscopes. This is the same conclusion
the present authors also came to.

Here we present some preliminary results of a study that increases the GC
AGB sample sizes substantially. With this new information we hope to
confirm or disprove the existence of these abundance differences.


\section{The AGB Samples -- Photometry \& Astrometry}

\begin{figure*}[ht]
\begin{center}
\resizebox{0.65\hsize}{!}{\includegraphics{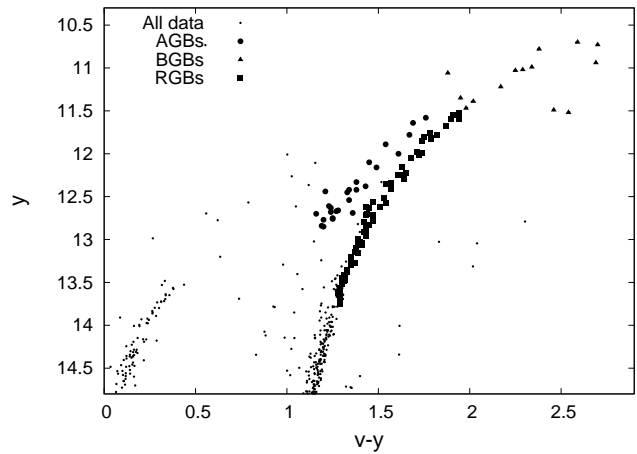}}
\end{center}
\caption{\footnotesize Example of star selections in the CMDs. This shows
  the selections for NGC 6752. Observational data are from \cite{GCL99}. 28
  AGBs are identified, 24 of which we now have spectra for.  BGBs are
  `bright giant branch' stars, defined because it is difficult to split the
  AGB and RGB in the brighter populations in this case. The blue
  `horizontal' branch can be seen at bottom left.}
\label{cmd1}
\end{figure*}

A vital ingredient in being able to find significant numbers of AGB stars
in globular clusters is having photometry good enough to separate the AGB
from the RGB.  Photometric observations have now reached such high accuracy
that it is becoming feasible to separate the AGB and RGB populations
reliably. During our literature search (\citealt{Paper1}) we came across
some very high-quality photometric studies, such as the study of M5 by
\citet{SB04}. Their set of observations is complete out to 8-10 arc
min. They also tabulate all their stars according to evolutionary status --
and find 105 AGB stars! This represents a 10-fold increase in sample
size. For the current study we have used the \citet{SB04} sample for M5,
plus a range of other CMDs, mostly from \cite{GCL99}. The Grundahl
CMDs are in the Str{\" o}mgren $uvby$ system (\citealt{STROM66}). In Figure
\ref{cmd1} we show an example of our CMD selection of AGB stars, for NGC
6752. In this case 28 AGB stars were identified and 24 were observed during
our observing run.

Using a multi-object spectroscopy instrument placed a further constraint on
the AGB samples. Very accurate astrometry was needed. This reduced some of
the sample sizes.  Also, due to crowding on the 2dF field plate, a few more
stars were lost from each sample. In all our AGB sample was reduced from
$\sim 500$ stars to $\sim 400$ stars, across 10 GCs: NGC 1851, NGC 288, NGC
362, NGC 6752, M2, M4, M5, M10, 47 Tuc, and Omega Cen.

In our GC sample we have included the outlier -- M5, which appears to
have a majority of CN-strong stars on its AGB, a fact that may cause
problems for some explanations of the (possible) phenomenon. We also
include 47 Tuc, which shows a mix of CN-weak and CN-strong (\citealt{M78}).

In addition to the AGB samples we also have selections of RGB stars, red HB
stars, and `bright giant branch' (BGB) stars (the brightest stars in the
CMDs, for which we couldn't split the RGB and AGB - see Figure
\ref{cmd1}). The RGB stars serve as checks, since these populations are
usually well studied in terms of CN, plus they have similar luminosities and
temperatures as the AGB stars. The HB stars may provide information on
which stars proceed to the second giant branch and which do not (assuming
the CN-weak AGB phenomenon is real).

In all we now have a database of $\sim 800$ stars at various stages of
evolution in 10 GCs with accurate photometry and astrometry.

\section{Observations}

The observing run consisted of 5 nights on the Anglo-Australian Telescope
(AAT). We used the multi-object spectroscope, AAOmega/2dF. On the blue arm
we used the 1700B grating, which gave a spectral coverage of 3755 to 4437
$\AA$, which includes the blue CN and CH bands. On the red arm we used the
2000R grating. The resolution of the spectra is $\sim 1.2 \AA$.  Near the
CN bands the S/N $\gtrsim 20$, rising to $\gtrsim 50$ at the CH bands.

\begin{figure*}[ht]
\resizebox{\hsize}{!}{\includegraphics[clip=false]{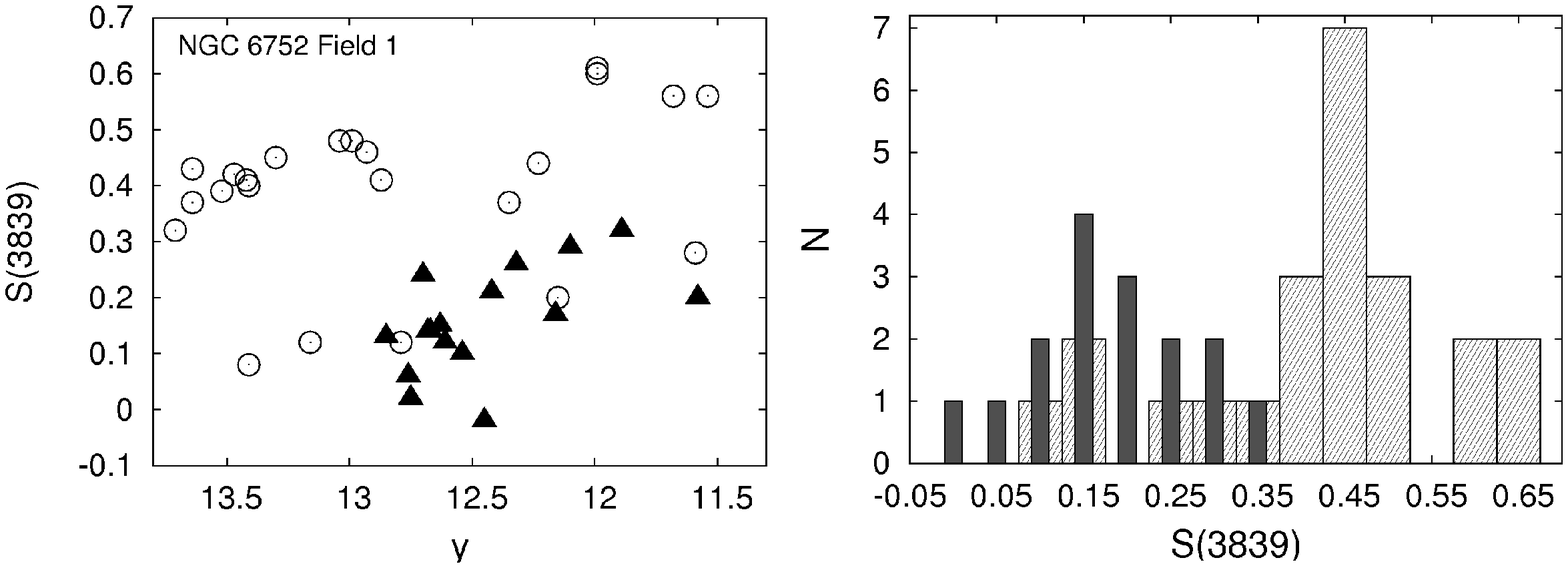}}
\caption{\footnotesize Preliminary results for NGC 6752 (Field 1 of the
  three 2dF fields observed). There are 16 AGB stars and 23 RGB
  stars. \textit{Left panel}: S3839 CN index versus $y$ magnitude. Open
  circles are RGB stars, filled triangles AGB stars. The trend with
  magnitude is due to stellar temperature effects. \textit{Right panel}:
  Histogram of the distribution of CN index values for both the the AGB
  sample (dark thin bars) and RGB sample (light bars). Bin size is
  0.05. The bimodality of the RGB and the monomodality of the AGB are
  clearly seen. $N$ is the raw number of stars.}
\label{6752results}
\end{figure*}
\begin{figure*}[ht]
\resizebox{\hsize}{!}{\includegraphics{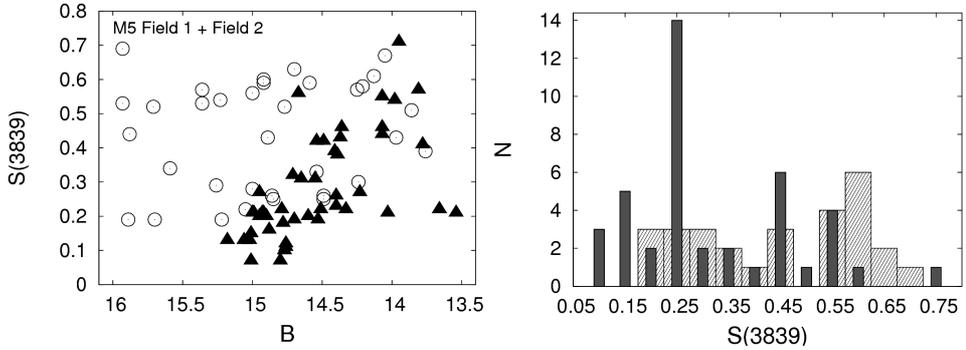}}
\caption{\footnotesize Preliminary results for M5 (both of the two 2dF
  fields observed are shown). There are 42 AGB stars and 28 RGB
  stars. \textit{Left panel}: S3839 CN index versus $B$ magnitude. Open
  circles are RGB stars, filled triangles AGB stars. The trend with
  magnitude (more easily seen in the RGB data) is due to stellar
  temperature effects. \textit{Right panel}: Histogram of the distribution
  of CN index values for both the the AGB sample (dark thin bars) and RGB
  sample (light bars). Bin size is 0.05. The bimodality of both the
  RGB and AGB can be seen, as can the strong bias towards CN-weak stars on
  the AGB. $N$ is the raw number of stars.}
\label{m5results}
\end{figure*}


\section{Preliminary Results}

Data reduction and analysis is still ongoing. We are using the 2dFdr
software provided by the AAO for initial reductions, then IRAF for
analysis. 

Although the spectra are only of moderate resolution it is sufficient to
check for cluster non-members (or binaries) using radial velocities. For
this we used the IRAF command $fxcor$. So far we have only found 1 or 2
stars per cluster that show strongly deviant velocities, which suggests
there is not much noise in our sample from non-members.

To quantify the CN band strengths in each star we use the S(3839) CN index
of \cite{NCF81} which compares a section of the CN bands with a neighboring
pseudo-continuum:
\begin{equation}
S(3839) = -2.5 \log \frac{\int_{3846}^{3883} \! I_\lambda \, d\lambda}{\int_{3883}^{3916} \! I_\lambda \, d\lambda}
\end{equation}

IRAF was used to measure the integrated fluxes. We discarded data with low
counts, effectively putting a lower limit on the S/N of $\sim 16$ at the CN
bands. In Figure \ref{6752results} we show some results for NGC 6752. This
is the cluster investigated by \cite{NCF81} that inspired the current
study. They reported that all of their 12 AGB stars were CN-weak, as
opposed to the RGB sample which showed bimodality. Our Figure \ref{6752results}
shows the same -- a clear bimodality in the RGB sample, whilst \textit{all}
the AGB stars are CN-weak! This preliminary result is based on one of the three
2dF fields, which includes 16 AGB stars and 23 RGB stars. We include a
histogram of this data in the right-hand panel of the figure. Indeed the
$100\%$ CN-weak AGB is very clear-cut, even without `de-trending' the data for
temperature effects on the CN band strengths.

In Figure \ref{m5results} we show the data for all of the M5 RGB and AGB
stars for which we have spectra. The 42 AGB and 28 RGB stars are a
combination of both the two 2dF fields that we observed. Here we can see
that the situation is not as clear-cut as in NGC 6752 -- the bimodality of
the RGB can be seen, however the AGB also presents a bimodal distribution
of CN-strong and CN-weak stars. However, contrary to the results of
\cite{SN93}, who report a CN-strong majority on the AGB, our sample shows a
definite CN-\textit{weak} majority on the AGB. The source of the
difference may be that the \cite{SN93} sample only contains 8 AGB stars.

Finally we note that our preliminary results for NGC 288 (not displayed
here) show a very similar picture to NGC 6752 -- a clear-cut bimodality on
the RGB and a totally CN-weak dominated AGB.


\section{Discussion}

The preliminary results of the current study strongly support the
unexpected phenomenon in which CN-rich stars seem to `disappear' between
the RGB and AGB, leaving CN-weak dominated AGBs. 

There is the possibility that the measurements of CN in AGB stars are
biased in some way, either by gravity or temperature differences as
compared to the RGB stars. However, as discussed by \cite{NCF81}, the fact
that we see stars at the same temperature on the RGB with varying CN band
strengths indicates that this is not affecting the results. The higher
luminosity of the AGB stars at the same temperatures would tend to increase
the CN band strengths, indicating that gravity also is not affecting the
results. Furthermore, in some of our observations we see CN-rich AGBs as
well as CN-weak AGBs in the same cluster, which again indicates that the
AGB measurements are not affected by their slightly different
properties. We do however plan to perform spectral synthesis calculations
to double-check. A benefit of the current study is that our spectra data
set is homogeneous (since all the spectra were taken on the same
instrument).

If we accept the above arguments and that the paucity (or total lack of)
CN-strong stars on the AGB is real, then we need to look at evolutionary
affects. There is however no known reason why the surface abundance of CN
(which is a proxy for N) should reduce when a star evolves from the RGB to
the AGB, at least in standard stellar evolution theory. Moreover, due to
the well-known `deep mixing' effect that increases the N abundances (at the
expense of C) in stars as they ascend the RGB (e.g. \citealt{S03}), this is
actually the opposite to what we would expect on the AGB -- we would
predict, if anything, more CN-strong stars on the AGB!

\cite{NCF81} proposed two possible explanations to explain the (apparent)
lack of CN-strong stars on the AGB:

\begin{enumerate}

	\item The two populations in NGC 6752 have different He abundances (they
	suggest $\bigtriangleup Y\sim0.05$). This may have come about through
	a merger of two proto-cluster clouds with differing chemical histories
	or through successive generations of stars (i.e. self-pollution). The
	He-rich material would also be N-rich. The He-rich stars would then
	evolve to populate the blue end of the HB -- and not ascend the AGB --
	leaving only CN-normal stars to evolve to the AGB. 

	\item Mixing in about half of the RGB stars pollutes their surfaces (increasing
	N) and also increases mass-loss rates, again leading to two separate
	mass groups on the HB. As before, the CN-strong, low mass group does
	not ascend the AGB.

\end{enumerate}

A constraint on the first explanation (for NGC 6752) is that about
\textit{half the mass of the cluster} must be polluted, as the number of
CN-strong and CN-normal stars is (usually) roughly equal. As Norris et
al. state, this is not a serious problem for the merger scenario, as the
merging clouds/proto-clusters may very well have had similar
masses. However, due to the constancy of Fe group elements, the chemical
histories of the two clouds/proto-clusters would have to have been
identical with respect to these heavy elements. This is more difficult to
explain since we require differing chemical histories for the light
elements.

In light of recent observations on the MS and SGBs of some clusters, the
second explanation by Norris et al. may require some clarification.  As N
appears to have a preformation source (as evidenced by MS observations),
the extra mixing is not required (although it does still exist). However,
the general suggestion that the differing compositions may affect mass-loss
rates and lead to different mass populations on the HB may be a valid
one. We are calculating stellar models that include this effect.

An important point apparent in the current results is that there appears to
be a variation in the number of CN-strong stars that `disappear' -- some
GCs show a total lack of CN-strong stars on their AGBs, whilst some show a
mix of CN-strong and CN-weak, although they are still dominated by CN-weak
stars.  Theories such as those of \cite{NCF81} will have to account
for this point also.

Finally, the substantial abundance differences between the GC RGBs and AGBs
reported here may reveal other clues to the GC abundance anomaly problems
(i.e. those of the heavier p-capture products - see \citealt{SIK00} for a
discussion), and even the GC formation problem. The rest of the results
from this study will be published soon, including a more in-depth analysis
and discussion.


\begin{acknowledgements}

The Authors wish to thank the Local and Scientific Organising Committees of
the Tenth Torino Workshop on Nucleosynthesis in AGB Stars, held at
Canterbury University in Christchurch, New Zealand. We also thank the
creators and maintainers of the very useful observational datasets and
software: 2MASS, GSC, 2dFdr, SIMBAD, IRAF, Aladin and VizieR. SWC
acknowledges the support of the Consejo Superior de Investigaciones
Cient\'{i}ficas (CSIC, Spain) JAE-DOC postdoctoral grant and the MICINN grant
AYA2007-66256. RJS acknowledges support from the Australian Research
Council's Discovery Projects funding scheme (project number DP0879472).

\end{acknowledgements}


\bibliographystyle{aa}

\end{document}